\documentclass[12pt]{spieman}  
\usepackage{amsmath,amsfonts,amssymb}
\usepackage{graphicx}
\usepackage{setspace}
\usepackage{tocloft}

\newcommand{\BZ}{{\rm BZ}}

\title{Quantized transport of two-dimensional
multifrequency solitons}

\author[a,b,*]{Fangwei Ye}
\author[c]{Yaroslav V. Kartashov}
\author[d]{Vladimir V. Konotop}

\affil[a]{School of Physics,
Chengdu University of Technology, Chengdu 610059, China}
\affil[b]{School of Physics and Astronomy, Shanghai Jiao Tong University, Shanghai 200240, China}
\affil[c]{Institute of Spectroscopy, Russian
Academy of Sciences, 108840 Troitsk, Moscow, Russia}
\affil[d]{Departamento de Física and Centro de
Física Teórica e Computacional, Faculdade de Ciências, Universidade de Lisboa, Campo Grande, Edifício C8, Lisboa}

\cftpagenumbersoff{figure}
\cftpagenumbersoff{table} 
\begin{document} 
\maketitle

\begin{abstract}
We consider Thouless pumping of two-dimensional quadratic solitons composed from coherently interacting fundamental frequency (FF) and second harmonic (SH) components propagating in a periodic $\chi^{(2)}$ material. The pumping is induced by two mutually sliding two-dimensional lattices defined by shallow transverse and longitudinal periodic refractive index modulation. Focusing on solitons in the semi-infinite gap, we find three distinct pumping scenarios: the absence of transport for small-amplitude solitons, non-quantized transport in a transient regime at intermediate amplitudes, and stable quantized transport for solitons with relatively large amplitudes. Different directions of the sliding velocity were investigated resulting in different trajectories of soliton center. The transition to the regime of quantized transport is found to depend strongly on phase mismatch between FF and SH waves that also influences stability properties of two-dimensional $\chi^{(2)}$ lattice solitons. We also show that pumping with longitudinal periods in the $x$ and $y$ directions, whose ratio approximates an irrational number, induces quantized transport whose direction rapidly converges, with increase of the accuracy of the approximation, to a limiting pumping direction determined by this irrational number and corresponding to truly incommensurate longitudinal periods.   
\end{abstract}


{\noindent \footnotesize\textbf{*}Fangwei Ye,  \linkable{fangweiye@sjtu.edu.cn} }

\begin{spacing}{2}   

\section{Introduction}

Quantized transport (also known as Thouless pumping) \cite{Thouless1983} is a fundamental physical phenomenon involving the quantized, typically adiabatic, transport of physical quantities, such as the total power of a light beam in the optical system considered in this work, across a periodic medium (a lattice) driven by slow periodic modulations of the lattice parameters. This effect has been discovered and experimentally confirmed in numerous linear systems of diverse physical origins, including atomic, optical, and acoustic platforms \cite{Lohse2016, Nakajima2016, Zilberberg2018, Lohse2018, Wang2022, Nakajima2021, Cerjan2020, Yang2024, Peng2025}. For recent overview of these results, we refer the reader to recent reviews~\cite{Citro2023, Szameit2024}.

In multiple optical settings as well as in studies of quantized transport of Bose-Einstein condensates, a nonlinearity is present as a natural phenomenon accompanying transport. In particular, nonlinearity enables the existence of solitons, thus raising the important problem of quantized transport of such spatially localized objects. The problem of soliton transport differs from conventional linear Thouless pumping in several important aspects. First, there is the constraint on the initial excitations: exact solitons possess a well-defined spectrum, which becomes narrow in the small-amplitude limit of envelope solitons, unlike in typical linear transport scenarios, where total band occupation is required. Second, nonlinearity unavoidably induces coupling between bands --  when finite band gaps exist -- while in the linear case different bands remain decoupled and support independent transport as long as they do not cross during adiabatic evolution. Third, nonlinearity leads to a strong, and in some cases complete, suppression of dispersion and may even lead to instabilities -- the phenomena absent in purely linear systems. Remarkably, despite these substantial differences, the quantized transport of solitons in cubic media (such as optical media with Kerr nonlinearity or Bose–Einstein condensates with two-body interactions) remains governed by the topological characteristics, specifically the Chern numbers, of the corresponding linear system as has been shown in several recent publications \cite{Nakagawa2018, Jurgensen2021, Fu2022a, Fu2022b, Jurgensen2022, Mostaan2022, Jurgensen2023, Lyu2024, Viebahn2024, Ye2025}.

Thouless pumping of solitons observed so far in periodic lattices only for ubiquitous cubic nonlinearities is a universal physical wave phenomenon that may have analogies in materials with other types of nonlinear response, but at the same time different type of nonlinear response may introduce completely new features into this phenomenon. For instance, solitons may form in periodic materials due to parametric interactions between different frequency components, as described in reviews \cite{Lederer2008, Kartashov2009, Chen2012, Kartashov2019}, while optical platforms provide unique testbed for combination of such interactions with dynamically changing periodic lattices. Such settings may allow demonstration of evolution regimes otherwise inaccessible in other physical systems, among which is the celebrated formation of solitons in quadratic optical media \cite{Torner2002, Buryak2002}. Such solitons have been studied theoretically and experimentally in different static lattice systems \cite{Lederer2008}, including incommensurate two-dimensional (2D) moir\'e lattices~\cite{Kartashov2021}.

Among the rich variety of solitons, multifrequency solitons occupy a special niche, not only because of their physical relevance~\cite{Torner2002,Buryak2002}, but also due to two seemingly incompatible features. First, the fundamental field (FF) and second harmonic (SH) components of such solitons propagate in different media -- though these may share similar topological properties -- and remain decoupled in the purely linear limit. As a result, their transport properties are governed by two distinct linear lattices. On the other hand, nonlinearity is essential for the very existence of multifrequency solitons, as it determines the coupling and joint evolution of the FF and SH components. Recently, Thouless pumping of multifrequency solitons in two mutually sliding lattices in quadratic medium was reported in~\cite{Kartashov2025}, but only in one-dimensional (1D) configuration. {\color{red} In this context we note that quadratic solitons in certain aspects are analogous to solitons in nonlocal Kerr media~\cite{Nikolov2003,Esbensen2012}. Thouless pumping in systems of both types was considered in~\cite{Kartashov2025} and~\cite{Ye2025}, respectively, showing some similarities of the soliton evolution.  }

At the same time, 2D configurations may bring much richer pumping scenarios and lead to challenges connected with the fact that 2D solitons are, in general, more fragile objects than their 1D counterparts. The reports on Thouless pumping of 2D solitons are scarce (see, e.g. \cite{Fu2022b}) and are so far limited exclusively to cubic nonlinear materials. 

The goal of the present paper is to extend the study of Thouless pumping of multifrequency solitons to two-dimensional media, where additional physical characteristics become important for the dynamics. Namely, now the pumping in linear limit of each component is characterized by two Chern numbers (thus, in total, the system is characterized by four Chern numbers). In linear regime such two-dimensional transport (of one-frequency beams) was experimentally observed in \cite{Wang2022}, but it was never studied for two coherently interacting fields. In such two-dimensional systems the nontrivial question about the connection between the direction of transport and sliding direction of two lattices arises. Furthermore 2D media are usually characterized by additional sources of instabilities, and thus the results obtained for 1D transport cannot be applied straightforwardly. We also mention that 2D lattices allow for principally new dynamical scenarios, like for example mutually incommensurate pumping in different spatial dimensions, that must be clearly distinguished from the observed in recent experiments 1D pumping in quasi-periodic media~\cite{Yang2024} and 1D bichromatic pumping with incommensurate frequencies~\cite{Peng2025}.

\section{The model}

We consider the propagation of a paraxial beam consisting of the FF component (frequency $\omega$) with the amplitude $\Psi_1$ and SH component (frequency $2\omega$) with the amplitude $\Psi_2$, governed by the dimensionless coupled equations
 \begin{align}
 	\label{main1}
 	i\frac{\partial \Psi_1}{\partial z}=-\frac{1}{2}\nabla^2 \Psi_1-V(\textbf{r},z)\Psi_1-\Psi_1^*\Psi_2  
 	\\
 	\label{main2}
 	i\frac{\partial \Psi_2}{\partial z}=-\frac{1}{4}\nabla^2 \Psi_2 +\beta\Psi_2-2V(\textbf{r},z)\Psi_2-\Psi_1^2 
\end{align}

Here $z$ is the propagation distance, $\textbf{r}=(x,y)$ is the vector in the transverse direction, $\nabla=(\partial_x, \partial_y)$, and $\beta$ is the phase mismatch between FF and SH components (we use the normalizations adopted in continuous models of Refs.~\cite{Kartashov2004, Kartashov2021}). The optical potential is considered separable: $V(\textbf{r},z)=V_x(x,z)+V_y(y,z)$ where each of the components consists of two sublattices, one of which slides with respect to static sublattice ($\xi=x,y$)
\begin{align}
    V_\xi(\xi,z)=p_{1\xi}\cos^2\left(\frac{\pi \xi}{d_{1\xi}}\right)+p_{2\xi}\cos^2\left(\frac{\pi (\xi-\alpha_\xi z)}{d_{2\xi}}\right) 
\label{eq2}
\end{align}
with small sliding angle $\alpha_{x,y}\ll 1$. Sublattices have amplitudes $p_{j\xi}$ and transverse periods $d_{j\xi}$ ($j=1,2$). In this work the consideration will be restricted only to lattices periodic in the transverse direction, i.e., all sublattice periods $d_{j\xi}$ are set commensurate. Respective lattice periods are denoted by $\Xi$, i.e. $\Xi=X,Y$. We note that previously the existence and properties of quadratic solitons in static 2D lattices (corresponding to $\alpha_x=\alpha_y=0$ in our case) were investigated in~\cite{Malomed2002, Xu2004, Xu2005, Susanto2007, Hang2009, Xu2009, Zhao2023}, for earlier works on their 1D counterparts and experiments see \cite{Bang1997, Peschel1998, Kobyakov1999, Sukhorukov2000, Iwanow2004}. Further we will explore two different situations, when sliding velocities $\alpha_{x,y}$ are selected such that pumping periods $Z_\xi= d_{2\xi}/\alpha_{\xi}$ are commensurate and therefore one can introduce the period of pumping for the total lattice, below denoted by $Z$ (if $Z_1$ and $Z_2$ are integers, then $Z=$lcm$(Z_1,Z_2)$), and when $\alpha_{x,y}$ are selected such that periods $Z_\xi$ are incommensurate and formally period of pumping is infinite.

\textcolor{red}{In dimensionless Eq. (\ref{main1}) and (\ref{main2}) the transverse coordinates $(x,y)$ are normalized to characteristic width $a$, the propagation distance $z$ is scaled to the diffraction length $k_1a^2$, $k_1=n_1(\omega)\omega/c$ and $k_2=n_2(2\omega)2\omega/c$ are the wavenumbers of the FF and SH components at frequencies $\omega$ and $2\omega$, the dimensionless amplitudes of FF and SH waves are given by $\Psi_{1,2}=(2\pi\omega^2\chi^{(2)}a^2/c^2)E_{1,2}$, where $E_{1,2}$ are dimensional field amplitudes, and $\chi^{(2)}$ is the involved second-order susceptibility tensor element, phase mismatch $\beta=(2k_1-k_2)k_1a^2$, and sublattice amplitudes $p_{j\xi}\sim \delta nk_1^2a^2/n_1$ are determined by the actual refractive index modulation depth $\delta n$. For example, in KTP crystals for characteristic width $a=10~\mu\textrm{m}$ and wavelength of FF wave of $\lambda=1~\mu \textrm{m}$, refractive index $n_1\approx1.74$, the propagation distance $z=1$ corresponds to $1.09~\textrm{mm}$, the dimensionless intensity $|\Psi_{1,2}|^2\sim 1$ corresponds to $1~\text{GW/cm}^2$ (typical lattice solitons are formed at intensities up to $10~\text{GW/cm}^2$ \cite{Torner2002,Buryak2002}), lattice periods $X,Y=1$ correspond to $10~\mu\textrm{m}$, lattice depth $p_{j\xi}\sim 3$ corresponds to the refractive index modulation depth $\delta n\sim 4.4\times10^{-4}$, the dimensionless value of phase mismatch $\beta\sim\pm 3$ corresponds to coherence length $|2k_1-k_2|^{-1}\sim 0.8 ~\textrm{mm}$.} 

Considering adiabatic character of pumping, we start with the (quasi-)stationary solutions of (\ref{main1}), (\ref{main2}) which are searched in the form  
\begin{align}
    \Psi_j=e^{ij\int_{0}^{z}b(\zeta)d\zeta} \psi_j(x,z), \quad  j=1,2 
\end{align}
where both $b(z)$ and $\psi_j(x,z)$ are slowly varying functions of $z$, such that in the leading order the derivatives $db(z)/dz$ and $\partial\psi_j(\textbf{r},z)/\partial z$ can be neglected, thus leading to the system, which by analogy with the 1D case~\cite{Kartashov2025} can be written in the form
\begin{align}
\label{eigen-nonlin}
    b\psi_1=-H_1\psi_1+\psi_1^*\psi_2, \quad b\psi_2=-H_2\psi_2+\frac{1}{2}\psi_1^2
\end{align}
Here $H_1=H_{\epsilon=1}$, $H_2=H_{\epsilon=2}$ and
\begin{align}
    H_\epsilon=-\frac{1}{2\epsilon^2}\nabla^2+\frac{\epsilon-1}{2}\beta-V(\textbf{r}, z)
\end{align}
This allows to obtain in the linear limit the decoupled linear eigenvalue problems for the FF ($\epsilon=1$) and SH ($\epsilon=2$) components,
\begin{align}
\label{eigen-lin}
    b_\nu^{\epsilon}(\textbf{k})\psi_{\nu \textbf{k}}^\epsilon(\textbf{r})=H_\epsilon\psi_{\nu \textbf{k}}^\epsilon(\textbf{r})
\end{align}
where $\textbf{k}\in [-K_x/2,K_x/2)\times [-K_y/2,K_y/2)$ is the Bloch wavevector in the reduced Brillouin zone (BZ), $\nu=1,2,...$ is the band number (the smallest index $\nu=1$ corresponds to the upper band), $K_x=\pi/X$ and $K_y=\pi/Y$ are the widths of the Brillouin zone associated with $x$ and $y$ directions, respectively. One can see that the Hamiltonians $H_1$ and $H_2$ are related by the continuous one-parametric deformation (alias homotopy), with $\epsilon$ being the homotopy parameter. This fact is important for establishing similarity of the topological characteristics of the Hamiltonians $H_{1,2}$.

\begin{figure*}
\includegraphics[width=\textwidth]{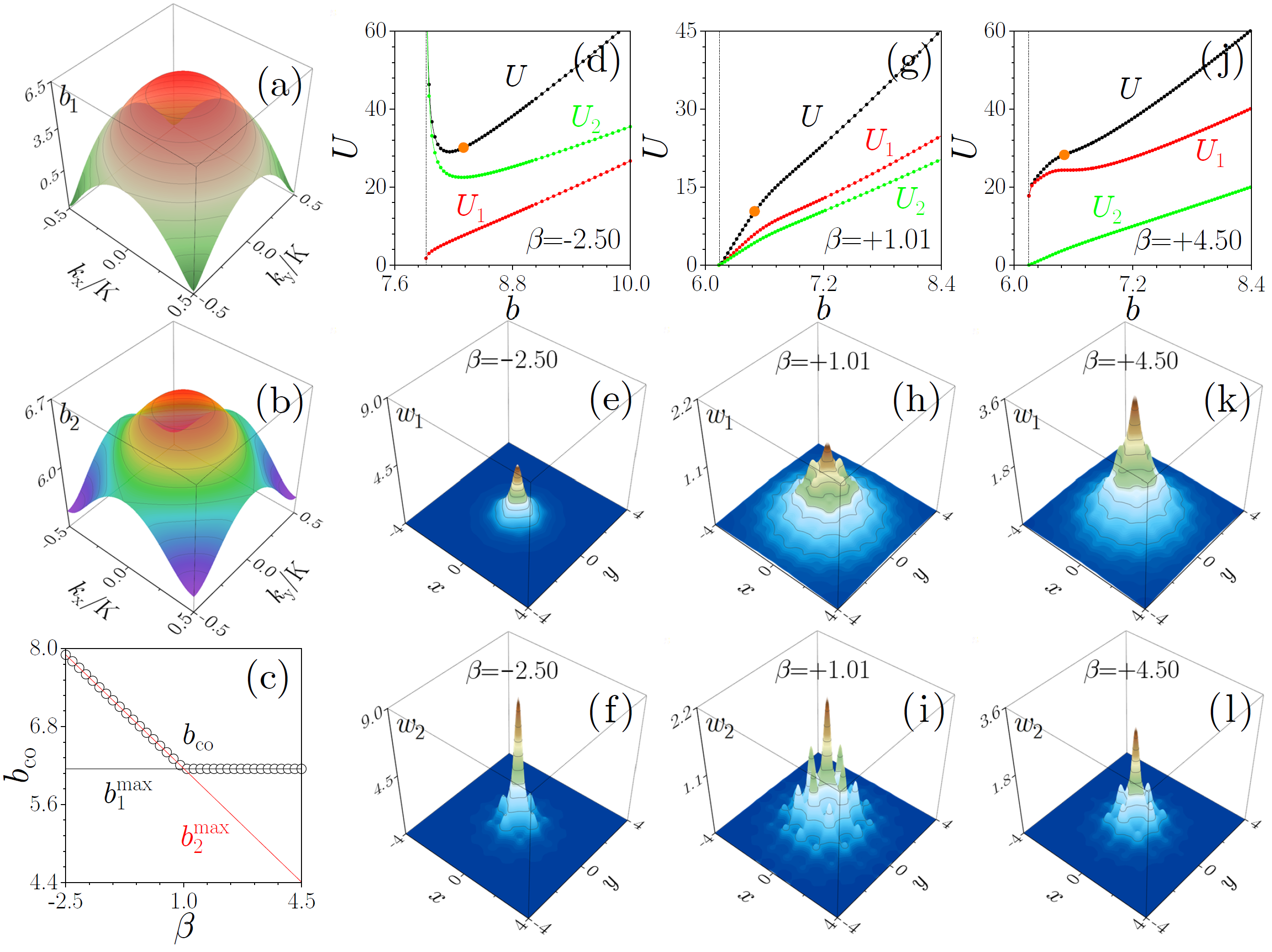}
\caption{\textcolor{red}{Linear spectrum of the lattice and families of lattice solitons.} Top band of static lattice for FF wave (a) and SH wave (b) at $\beta=0$. (c) Cutoff $b_\textrm{co}$ for 2D lattice solitons from semi‐infinite gap versus phase mismatch $\beta$. The dependencies $b_{1,2}^\textrm{max}(\beta)$ illustrate upper edges of the spectra of $H_{1,2}$. Total power and powers in FF and SH components versus propagation constant $b$ at $\beta=-2.50<\beta_\textrm{cr}$ (d), $\beta=+1.01\approx \beta_\textrm{cr}$ (g), and $\beta=+4.50>\beta_\textrm{cr}$ (j). Panels (e),(f) show profiles of soliton components corresponding to orange dot in (d), panels (h),(i) show profiles corresponding to orange dot in (g), while panels (k),(l) show profile corresponding to the orange dot in (j). Here and in all figures below $p_{1\xi} =p_{2\xi} =3.0$, $d_{1\xi} =0.5$, $d_{2\xi} =1.0$ (where $\xi=x,y$).}
\label{figure1}
\end{figure*}

The upper allowed bands \textcolor{red}{(i.e. the dependencies of linear eigenvalues $b_\nu$ on Bloch momentum $\textbf{k}$)} in spectra of $H_1$ and $H_2$ at $z=0$ are illustrated in Figures \ref{figure1}(a) and (b). \textcolor{red}{Corresponding dependencies $b_\nu(\textbf{k})$ for FF and SH waves were obtained numerically by solving linear eigenvalue problem (\ref{eigen-lin}) using plane-wave expansion method.} Notice that the upper band in spectrum of SH wave is substantially narrower than the upper band in the spectrum of FF wave. We denote the upper edge of the top band in each component as $b_\epsilon^{\rm max}=\max_{\textbf{k}\in\BZ}b_{\nu=1}^\epsilon(\textbf{k})$ \textcolor{red}{(i.e. as a maximum over all $\textbf{k}$ values belonging to the first two-dimensional Brillouin zone for a given wave component)}. In these notations, stationary lattice soliton solutions can exist in the \textit{total} semi-infinite gap, i.e., above the cut-off $b_{\rm co}=\max\{b_1^{\rm max},b_2^{\rm max}\}$, that is shown with open circles in Figure~\ref{figure1}(c) for $z=0$. It should be stressed that while $b_1^\textrm{max}$ does not depend on phase mismatch $\beta$ because this parameter does not enter corresponding Hamiltonian $H_1$ for $\epsilon=1$, the $b_2^\textrm{max}$ value decreases linearly with $\beta$ as shown with red line in Figure~\ref{figure1}(c). The curves indicating upper gap edges in FF and SH components cross at the critical value of phase mismatch $\beta_\textrm{cr} \approx 1.01$ (this critical value depends on the lattice shape and depth), determining the specific form of $b_\textrm{co}(\beta)$ dependence. Strictly speaking, $b_{\rm co}$ is also a function of the propagation coordinate $z$. Thus, pumping of 2D multicolor solitons is possible if propagation constant of soliton $b>\max_{z\in Z}b_{\rm co}(z)$.   

We now consider properties of 2D solitons at the input $\psi_j(\textbf{r},z=0)=w_j(\textbf{r})$. \textcolor{red}{Soliton profiles were obtained from nonlinear Eq. (\ref{eigen-nonlin}) using Newton iterations method. By varying propagation constant $b$ at fixed phase mismatch $\beta$ one can obtain the entire soliton families. Typically, in Newton method we used transverse windows of $41X\times41Y$ and $801\times801$ transverse points.} Below such solitons will be used as initial conditions for illustration of Thouless pumping. There are three different situations \cite{Moreira2012, Moreira2013, Kartashov2021}, which are outlined in Figure~\ref{figure1}(c). At $\beta<\beta_\textrm{cr}$, when $b_2^\textrm{max}>b_1^\textrm{max}$ the solitons do not have linear limit requiring a finite total power $U=U_1+U_2=\int(|\Psi_1|^2+|\Psi_2|^2)d\textbf{r}$ for their excitation [see the families of solutions depicted in Figure~\ref{figure1}(d) and representative profiles of two soliton components shown in panels (e) and (f)]. In this case the SH wave carries more power than the FF wave [see green and red lines in Figure~\ref{figure1}(d)]. The energy imbalance infinitely grows when propagation constant approaches the cutoff value $b\to b_{\rm co}$. The situation is opposite at $\beta>\beta_\textrm{cr}$, when $b_1^\textrm{max}>b_2^\textrm{max}$ [see the families in panel (j) and illustration of profiles of soliton components in panels (k) and (l) in Figure~\ref{figure1}]. Now the FF wave carries larger power than SH wave, although at $b\to b_{\rm co}$ the total power $U$ remains finite. The two cases mentioned above are separated by the specific third one realized at $\beta=\beta_\textrm{cr}$, where $b_{\rm co}=b_{1}^{\rm max}=b_{2}^{\rm max}$ [see soliton family shown in Figure~\ref{figure1}(g) and examples of profiles of soliton components in panels (h) and (i)]. In this critical case the FF and SH waves carry nearly equal power. Such solitons do bifurcate from the linear limit, since both $U_{1,2}\to 0$ as $b\to b_{\rm co}$. In the cutoff all solitons strongly expand across the lattice, while sufficiently far from the cutoff they contract practically to a single transverse lattice period. 

\section{Commensurate pumping}

\subsection{Solitons bifurcating from the linear limit}

We start the discussion of Thouless pumping of 2D quadratic solitons with this last case allowing for the linear limit, assuming that at $z=0$ a solitons with a given propagation constant $b$ is launched into dynamically varying lattice (\ref{eq2}) due to mutual sliding of two sublattices. \textcolor{red}{The dynamics of propagation of solitons in the dynamically varying lattice governed by Eqs. (\ref{main1}) and (\ref{main2}) was modeled using split-step fast-Fourier transform method.} To quantify the soliton displacement, we use the coordinates of its center of mass $(x_c(z),y_c(z))$ computed as $\xi_c(z)=U^{-1}\int \xi (|\Psi_1|^2+|\Psi_2|^2)d\xi$. First, we consider the case when second sublattice slides with respect to the first one along the diagonal direction, i.e. $\alpha_x=\alpha_y$.

\begin{figure*}
\includegraphics[width=\textwidth]{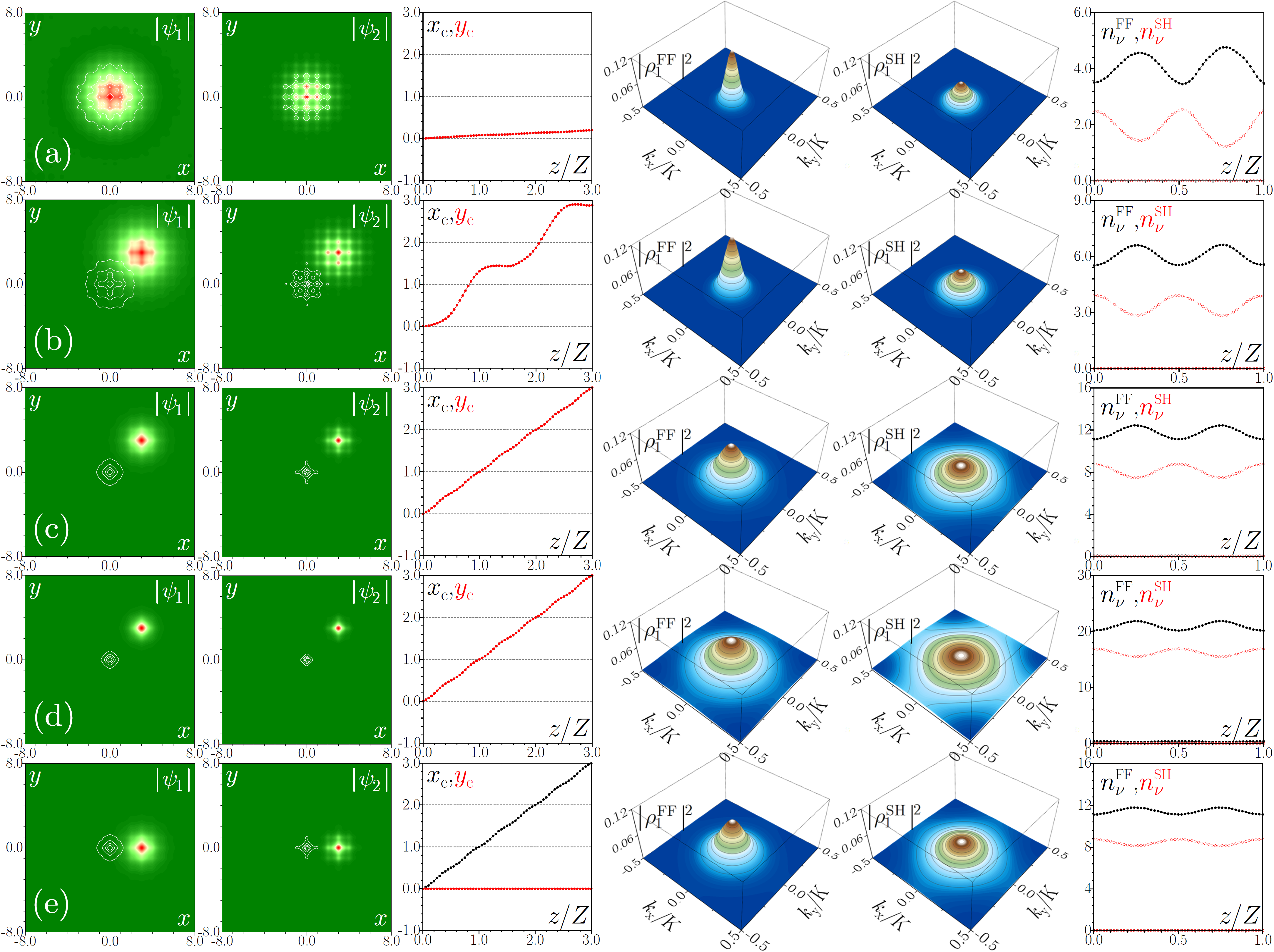}
\caption{\textcolor{red}{Dynamics of topological pumping of 2D quadratic solitons and the role of band filling.} Diagonal pumping of two‐dimensional quadratic solitons with propagation constants $b=6.35$ (a), $6.47$ (b), $7.00$ (c), $8.00$ (d) at $\alpha_x =\alpha_y =0.01$. First and second columns show input ($z =0$, white contour) and output ($z =3Z$ , red‐green color map) modulus distributions of the FF and SH components. Third column shows $x‐$ and $y‐$coordinates of soliton center of mass vs propagation distance on three pumping cycles. Fourth and fifth columns show, respectively, projections of the FF and SH components of soliton on the first band of dynamical optical lattice at $z =0$. Sixth column shows total populations of the bands in FF wave (black dots) and SH wave (red dots) vs distance $z$ on one pumping cycle. Bottom row (e) illustrates pumping of quadratic soliton with propagation constant $b=7.00$ along the $x‐$axis at $\alpha_x =0.01$, $\alpha_y =0$. In all cases $\beta=+1.01$.}
\label{figure2}
\end{figure*}

For a small-amplitude and broad soliton with $b=6.35$ taken sufficiently close to cutoff $b_{\rm co}$, slow sliding $\boldsymbol{\alpha}$ does not lead to quantized transport, as it is illustrated in the upper row of Figure~\ref{figure2} obtained for $\boldsymbol{\alpha}=(0.01,0.01)$. Only a small shift of the center of mass can be observed, as shown in the third column of the Figure (recall that the longitudinal lattice period for such sliding velocity is large, more specifically $Z=10^2$, and we propagate soliton in Figure \ref{figure2} up to $z=3Z$ to make even small displacements obvious). Suppression of transport for solitons at small amplitudes have been observed also for two-dimensional Kerr solitons~\cite{Fu2022b}, for 1D quadratic solitons~\cite{Kartashov2025}, as well as for 1D solitons in nonlocal media~\cite{Ye2025}. However already at slightly larger propagation constant $b=6.47$ in line (b) of Figure~\ref{figure2} one observes transport of soliton, which is not quantized yet. Moreover during the first cycle of pumping the shifts of center of mass of soliton in the $x$ and $y$ directions are larger than the lattice periods $X$ and $Y$, meaning that soliton moves with the average velocity larger than the pumping velocity $|\boldsymbol{\alpha}|$. During the second cycle the pumping slows down, while during the next third cycle it accelerates again. Further increase of propagation constant (power) of input soliton leads to established quantized transport that is illustrated in line (c) of Figure~\ref{figure2} (see the dynamics of center of mass in third column that exhibits in this regime quantized displacement by one $X$ and $Y$ period after each pumping cycle $Z$).

This behavior is tightly connected with projections of corresponding input solitons on the bands of the lattice. Such projections on the band with index $\nu$ at distance $z$ can be calculated as
\begin{align}
    \rho_\nu^{\rm FF,SH}(\boldsymbol{k})=\int \bar{u}_{\nu\boldsymbol{k}}^{1,2}(\boldsymbol{r})\Psi_{1,2}(\boldsymbol{r},z) d\boldsymbol{r},
\end{align}
where $u_{\nu\boldsymbol{k}}^\epsilon=\psi_{\nu\boldsymbol{k}}^\epsilon e^{-i\boldsymbol{k}\cdot\boldsymbol{r}}$ is the periodic part of the respective Bloch function in the spectrum of FF or SH wave calculated at a given distance $z$ and corresponding to Bloch momentum $\boldsymbol{k}$ within first Brillouin zone, $\Psi_{1,2}(\boldsymbol{r},z)$ is the field distribution of the FF and SH waves at the same distance $z$, and the overbar denotes the complex conjugation. The quantities $|\rho_\nu^{\rm FF}|^2$ and $|\rho_\nu^{\rm SH}|^2$ characterize soliton spectral densities. It should be mentioned that for our parameters the projections $ \rho_\nu^{\rm FF,SH}(\boldsymbol{k})$ on the first band $\nu=1$ always strongly dominate (because corresponding solitons originate from the semi-infinite gap laying above this band) and we further show only them in the fourth and fifth columns of Figure \ref{figure2}. The projections are shown at the input, at $z=0$ -- they change during pumping process (primarily due to power exchange between FF and SH components induced by lattice $z$-variation), but not qualitatively. One can observe that low-power broad solitons feature narrow spectral densities [see fourth and fifth columns in Figure \ref{figure2}(a),(b)], but they become much wider for strongly localized soliton states that exhibit robust and quantized Thouless pumping [see fourth and fifth columns in Figure \ref{figure2}(c),(d)]. Also, due to overall stronger spatial localization of SH wave, its spectral density distribution is always wider than that for FF wave.

Since in the linear limit FF and SH waves become decoupled, the Bloch functions for our potential factorize: $u_{\nu\boldsymbol{k}}^\epsilon(\boldsymbol{r})=u_{\nu_x k_x}^{x, \epsilon}(x) u_{\nu_y k_y}^{y, \epsilon}(y)$, where $\nu_{x,y}$ are the band indices in the respective directions, and one can use for characterization of topological properties of the bands the conventional 1D space-time Chern indices for the $x$ and $y$ lattices 
\begin{align}
\label{Chern}
C_{\nu_\xi }^{\xi,\epsilon}= \frac{1}{\pi }\text{Im}\int_{0}^{Z_\xi}dz\int_{-\pi/\Xi}^{\pi/\Xi}dk\langle  \partial_{k} u_{\nu_\xi k}^{\xi,\epsilon},  \partial_z u_{\nu_\xi k}^{\xi,\epsilon}\rangle .
\end{align}
Here $\langle f,  g \rangle:=\int_{0}^{\pi} \bar{f}(\xi)g(\xi)d\xi$, and the integration is carried over the corresponding longitudinal period $Z_\xi$ for the $\xi$-direction, and over the corresponding Brillouin zone for this direction. For the lattice used in Figure~\ref{figure1} and Figure~\ref{figure2} one computes four linear Chern indices $C_{1}^{\xi,\epsilon}=1$ for $\xi=x,y$ and $\epsilon=1,2$ for the top band $\nu=1$. This means that the quantized one-cycle displacements of solitons along the $x$ and $y$ axes observed in third column of Figure \ref{figure2}(c) and (d) are in exact agreement with the \textit{linear} theory of Thouless pumping. The comparison of spectral densities in panels (a),(b) and in panels (c),(d) of Figure \ref{figure2} also clearly reveals the importance of the sufficiently complete filling of the band that is not granted in optical systems. We stress that even further increase of propagation constant (and, hence, power) of input soliton in Figure~\ref{figure2}(d) keeps transport quantized and does not lead to its arrest.

In all cases where the quantized transport is observed, we notice significantly weaker power exchange between the FF and SH waves in comparison with the cases, where transport is not quantized. This is evident from the sixth column of Figure \ref{figure2}, where we show the evolution of the total band populations
\begin{align}
n_\nu^{\rm FF,SH}(z)=\int |\rho_\nu^\textrm{FF,SH}(\boldsymbol{k},z)|^2 d\boldsymbol{k},
\end{align}
with propagation distance $z$. Notice that in this figure we show total populations of the top five bands, but the population of band $\nu=1$ always dominates. From these dependencies one can estimate that only $\sim 6\%$ of the total power is transferred between the FF and SH waves in  Figures \ref{figure2}(c),(d), while in Figure \ref{figure2}(a) the power transfer increases to $\sim 13.3\%$.

Finally, in Figure \ref{figure2}(e) we provide an example of 2D soliton pumping dynamics in the case, when sliding of the second sublattice occurs only along the $x$-direction, i.e. for $\alpha_x=0.01$ and $\alpha_y=0$. This situation is unique for 2D geometry and it illustrates that in this setting, the possible pumping scenarios can be much richer in comparison with those in the 1D geometry. In this particular case $C_{1}^{y,\epsilon}=0$, while the indices $C_{1}^{x,\epsilon}=1$. As a result, the solitons with sufficiently large power launched into such lattice exhibit quantized transport only in the $x$-direction, i.e. in the direction of sliding. Notice that in this case the oscillations of total band populations with distance $z$ [Figure \ref{figure2}(e), sixth column] are smaller than oscillations of populations for the same soliton pumped along the diagonal [Figure \ref{figure2}(c), sixth column].

\subsection{Solitons having no linear limit}

So far we considered the case of solitons at $\beta=\beta_\textrm{cr}$ having linear limit, whose power $U$ vanishes as $b\to b_\textrm{co}$ [Figure~\ref{figure1} (g),(h),(i)]. Now we turn to the case, where the power of solitons do not vanish as propagation constant approaches the cutoff and analyze the one-cycle displacement of such solitons as a function of the propagation constant, and hence, of the amplitude of the soliton. 

\begin{figure}
\includegraphics[width=\columnwidth]{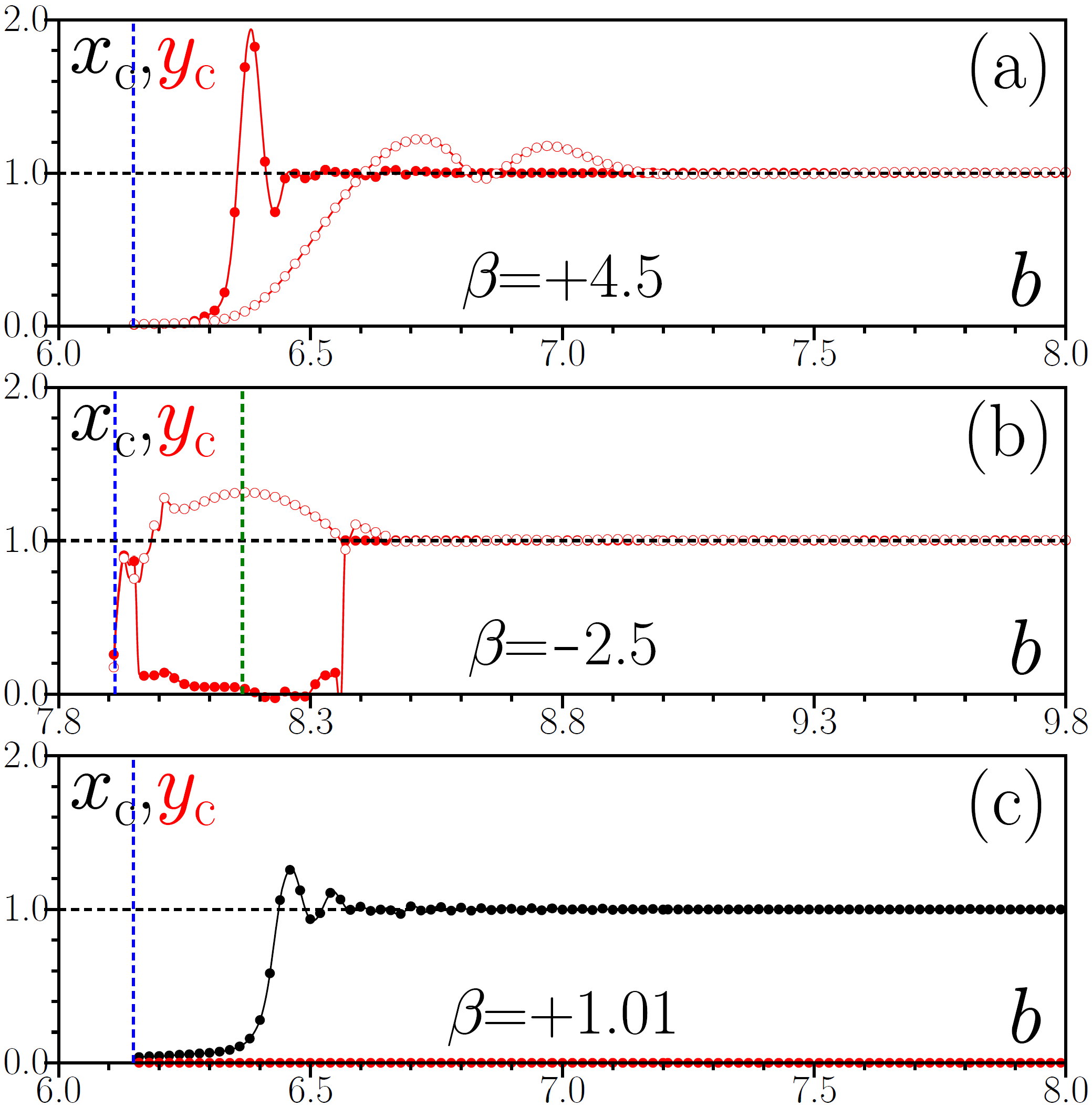}
\caption{\textcolor{red}{Soliton center of mass shift vs propagation constant.} Coordinates of the soliton center of mass at $z=Z$ versus propagation constant for $\beta=+4.5$ (a) and  $\beta=-2.5$ (b) upon diagonal pumping with \textcolor{red}{$\alpha_x=\alpha_y=0.01$ (solid dots) and $\alpha_x=\alpha_y=0.05$ (open dots).} In this case $x_c$ and $y_c$ coincide. Vertical blue dashed lines indicate cutoff for soliton existence. Vertical greed dashed line in (b) corresponds to $b$ value at which total soliton power $U$ reaches its minimal value. (c) Coordinates $x_c$, $y_c$ of the soliton center of mass at $z=Z$ versus propagation constant for $\beta=+1.01$  upon pumping along the  $x-$axis at  $\alpha_x=0.01$, $\alpha_y=0$.}
\label{figure3}
\end{figure}

The representative dependence of the one-cycle soliton displacement on $b$ upon diagonal pumping (i.e. for $\alpha_x=\alpha_y=0.01$) for $\beta=+4.5>\beta_\textrm{cr}$ is presented in Figure \ref{figure3}(a)\textcolor{red}{(see curve with solid dots)}. Corresponding soliton family is presented in Figure \ref{figure1}(j). One can see that at relatively small detuning $b-b_{\rm co}\lesssim 0.2$ (the cutoff is indicated with blue dashed line), the sliding lattice does not lead to any displacement of low-amplitude solitons. As soliton amplitude increases, within a finite transition region in $b$, one observes transport, which is however not quantized and at certain values of propagation constant can be even "faster" than the "velocity" determined by sliding (see e.g. the maximum of the curve at $b\approx 6.4$). After this transition region, where the displacement $x_c,y_c$ exhibits decaying oscillations, one observes the transition to the regime of quantized soliton transport that in our case extends at least up to the values of $b>30$, i.e. at least within this range of propagation constant the transport of 2D solitons in $\chi^{(2)}$ medium is not arrested by the nonlinearity. We found very similar dependencies of the one-cycle displacement on $b$ for all values of phase mismatch $\beta>\beta_\textrm{cr}$, including the critical case $\beta=\beta_\textrm{cr}$ [see Figure \ref{figure3}(c) where such dependence is depicted for the case of pumping along the $x$-axis only -- it also clearly shows the transition between regimes without transport and with quantized transport and only oscillations of $x_c$ displacement in transition region are slightly smaller]. 

Quite a different dependence of the one-cycle displacement on propagation constant is observed for solitons, whose power $U$ rapidly grows as $b\to b_{\rm co}$ that occurs for phase mismatch values $\beta<\beta_\textrm{cr}$. The exemplary soliton family in this regime is presented in Figure \ref{figure1}(d). Corresponding dependence of soliton displacement on $b$ for $\beta=-2.5<\beta_\textrm{cr}$ \textcolor{red}{and small sliding velocity $\alpha_x=\alpha_y=0.01$} is show in Figure~\ref{figure3} (b) \textcolor{red}{with solid dots}. One observes that transition to quantized transport with increase of $b$ is very sharp, without oscillations and it is connected with stability properties of solitons in dynamically changing refractive index landscape. Namely, at $b<8.36$ the solitons are unstable, they quickly loose their original structure and diffract that leads to practically no displacement of the center. Only very close to cutoff the displacement may become nonzero. At $b>8.36$, when soliton remains stable in the course of propagation in $z$-periodic lattice, it does exhibit quantized transport. Remarkably, soliton stabilization occurs for propagation constant value exceeding the value corresponding to $dU/db=0$ point [that is indicated with green dashed line in Figure \ref{figure3}(b)]. In this case we also do not observe the arrest of transport with increase of the propagation constant (that contrasts with pumping of one-component 2D solitons in Kerr media~\cite{Fu2022b}). 

\textcolor{red}{To address the impact of sliding velocity on soliton center displacement, specifically in the context of the validity of the adiabaticity conditions in our setting, we calculated the soliton displacement for the case of substantially larger sliding velocity $\alpha_x=\alpha_y=0.05$, see open dots in Fig.~\ref{figure3}(a) and \ref{figure3}(b). As one can see, the transition to quantized transport occurs even in this case, even though it happens at somewhat larger propagation constant values as compared to the $\alpha_x=\alpha_y=0.01$ case. Moreover, similar transition to quantized transport is observed even for sliding velocities as large as $\alpha_x=\alpha_y=0.20$ (that may be considered as corresponding to deeply non-adiabatic regime, since in this case the longitudinal period is only $Z=5$). This is an indication that in this system one can realize fast topological pump, and there is no clear crossover between adiabatic and non-adiabatic regimes in terms of sliding velocity, since solitons with sufficiently large propagation constants still can exhibit quantized displacement, even at large $\alpha_{x,y}$. This indicates that sliding velocities $\alpha_{x,y}\sim0.01$ can be considered as satisfying conditions of adiabaticity.}

\begin{figure}
\centering
\includegraphics[width=0.75\columnwidth]{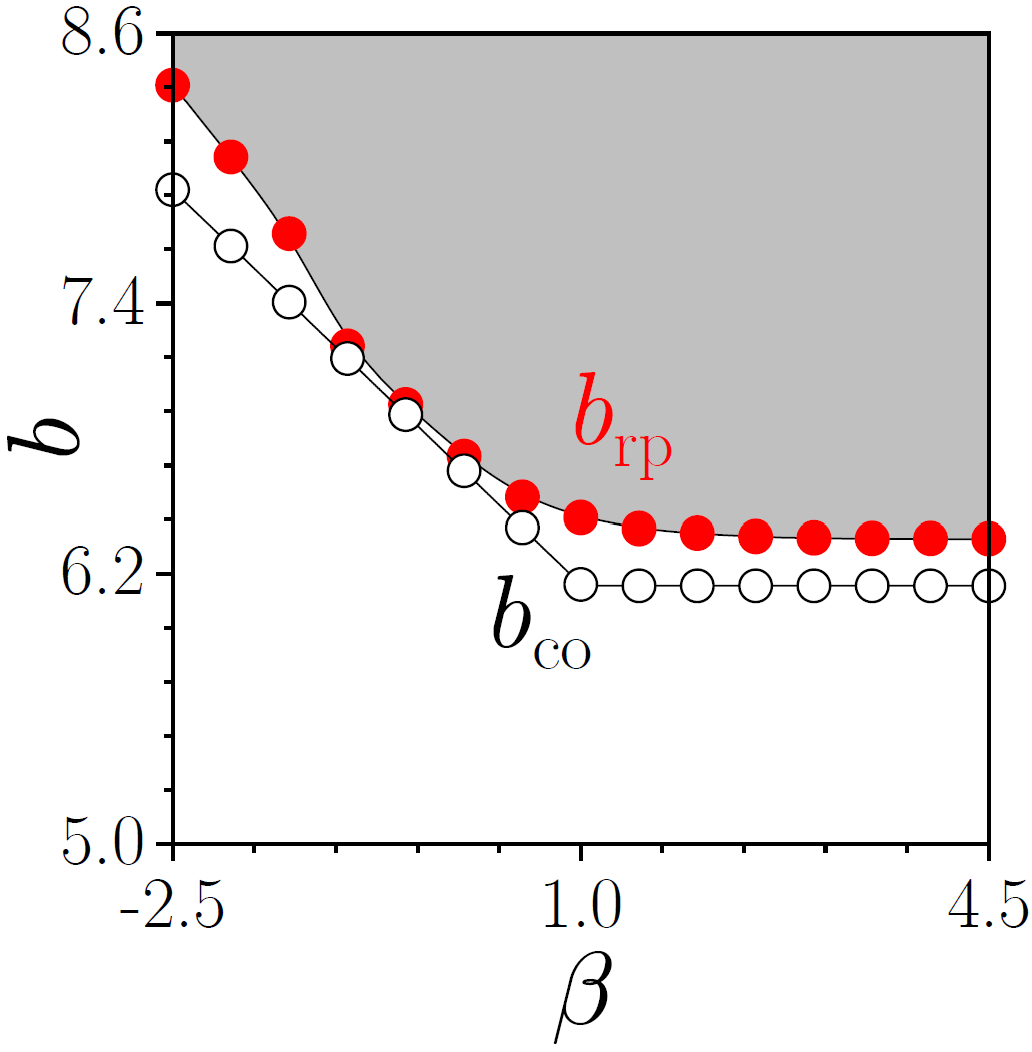}
\caption{\textcolor{red}{Domain of robust pumping of solitons on the $(\beta,b)$ plane. Line with open circles shows cutoff $b_\textrm{co}$ for soliton existence vs phase mismatch $\beta$. Line with red circles shows propagation constant $b_\textrm{rp}$ above which (shaded domain) robust pumping of solitons occurs at $\alpha_x=\alpha_y=0.01$.}}
\label{figure4}
\end{figure}

\textcolor{red}{As one can see from Fig. \ref{figure3}, robust pumping of solitons takes place above certain propagation constant value $b_\textrm{rp}$ and is possible in both $\beta>\beta_\textrm{cr}$ and $\beta<\beta_\textrm{cr}$ domains. To obtain the dependence $b_\textrm{rp}(\beta)$, we simulated pumping dynamics for all values of propagation constants $30>b>b_\textrm{co}$ with step $\delta \beta=0.5$ in phase mismatch at sliding velocity $\alpha_x=\alpha_y=0.01$. Corresponding dependencies of soliton center of mass shift on propagation constant $b$ for all values of $\beta$ can be found in Supplementary Figure S1. The value $b_\textrm{rp}$, at which the transition to robust pumping occurs, was defined as a value of $b$ at which soliton shift $x_c,y_c$ reaches transverse period $X,Y$ for the first time when $b$ increases from $b_\textrm{co}$. Robust pumping therefore occurs in shaded domain in Fig. \ref{figure4}. The width of the transition region $b_\textrm{rp}-b_\textrm{co}$ changes non-monotonically with $\beta$  and is minimal around $\beta=-1.0$. For large positive $\beta$ values this region remains practically unchanged. Notice that in the transition region solitons typically substantially spread at $\beta<\beta_\textrm{cr}$ or show only weak radiation at $\beta>\beta_\textrm{cr}$ on one pumping cycle $Z$.}

\section{Incommensurate pumping}

Incommensurate topological pumping can be implemented in a variety of settings. For example, one can consider pumping when one or both potentials $V_{\xi}$ are incommensurate along the $\xi$ direction (thus generalizing the observations of~\cite{Yang2024} to two dimensions). Another possibility is to consider bichromatic incommensurate pumping, when $V(\textbf{r},z)$ is a quasi-periodic function of $z$ (the respective 1D transport was recently reported in~\cite{Peng2025}). Remarkably, in 2D geometry one can also consider the transport which is the "conventional" periodic pumping in each of the directions, but having incommensurate transverse ($X$ and $Y$) or longitudinal ($Z_x$ and $Z_y$) periods. Here we explore this last setting, which can be viewed as an example of 2D incommensurate bichromatic pumping.

Such incommensurate pumping requires formally an infinite medium as soon as no finite longitudinal period is available. This raises a question about the very meaning of the quantization of transport upon such pumping, as soon as it cannot be rigorously implemented experimentally or numerically. The answer is positive, and resides in the use of the best rational approximations (BRA) for irrational numbers~\cite{Khinchin}. More specifically, assuming the same transverse periods $X=Y$, one can consider the relation of sliding velocities of sublattices $\alpha_y/\alpha_x=\gamma$, where $\gamma$ is an irrational number, which can be represented in a form of a continued fraction $\gamma=[a_0;a_1,a_2,...]$. Then, the number $p_n/q_n$, where $p_n$ and $q_n$ are coprime integers, obtained as the $n$-th order convergent: $p_n/q_n=[a_0;a_1,...,a_n]$ is the $n$-th order of BRA of $\gamma$. At $n\to\infty$ one has $p_n/q_n\to\gamma$.

Now we can formulate the main approach to study bichromatic 2D incommensurate topological transport: We investigate the spatial displacement $(X_n,Y_n)$ of the soliton center obtained in the $n$-th BRA as a function of growing order of approximation $n$ (instead of challenging the mathematical limit $n\to\infty$). One can then observe that if the quantized transport occurs, the ratio of displacements $Y_n/X_n$ over one pumping cycle converges very fast to the value
\begin{align}
    \frac{Y_n}{X_n}=\frac{p_n}{q_n}\to \gamma~~\textrm{as}~n\to\infty
\label{displacement}
\end{align}
This mechanism of the convergence of the 2D bichromatic pumping to incommensurate one is, however, very different from results of 1D experiments reported in Ref.~\cite{Peng2025}. Indeed, in the linear limit the $x$- and $y$-lattices remain longitudinally periodic with longitudinal periods related as $Z_x/Z_y=p_n/q_n$. This means that during one pumping cycle, whose period can be denoted as $Z_n$, the $x$- and $y$-lattices will undergo transformation corresponding to $q_n$ and $p_n$ longitudinal periods, respectively, so that total period of pumping (period of evolution of the entire 2D structure) will be equal to $Z_n=q_nZ_x=p_nZ_y$. Therefore, during one pumping period $Z_n$ the displacement of soliton center of mass in the regime of quantized transport will be equal to $X_n=q_nC_1^{x \epsilon}$ and $Y_n=p_nC_1^{y \epsilon}$. Since in the case at hand  $C_1^{\xi \epsilon}=1$, one expects to obtain the ratio of displacements of soliton center of mass given by (\ref{displacement}).

This rapid convergence to the incommensurate pumping is illustrated in Figure~\ref{figure4}, where we show in panel (a) the trajectories of the center of mass of 2D soliton with $b=7$ at $\beta=1.01$ [see corresponding soliton family in Figure \ref{figure1}(g)] in the transverse plane $(x,y)$ during one pumping cycle $z=Z_n$ for sliding velocities $\alpha_x=0.01$ and $\alpha_y=(p_n/q_n)\alpha_x$ corresponding to BRAs of progressively increasing order $n$ of the golden ratio $\gamma=\varphi$ where
\begin{align}
    \varphi=\frac{\sqrt{5}+1}{2}=\left[1; 2, \frac{3}{2},\frac{5}{3},\frac{8}{5},\frac{13}{8},\cdots \right]
\end{align}
that also determines the ratio of periods $Z_x/Z_y=p_n/q_n$. The evolution of the coordinates $x_c(z)$ and $y_c(z)$ of the center of mass of soliton with distance $z$, as well as initial and final soliton positions for different BRAs are presented in panels (b)-(e). One can observe that the trajectory quickly approaches (on average) the trajectory $y=\varphi x$. Since with increase of BRA order $n$ total longitudinal period $Z_n=q_nZ_x=p_nZ_y$ rapidly increases, total soliton displacement over one pumping cycle increases as well. While the respective displacement may seem to trivially reproduce Thouless predictions, the evolution of the well localized solitons shown in lower panels of Figure~\ref{figure4}(b)-(e) illustrates the dynamics far from being trivial. Indeed, in purely linear case, where FF and SH waves are not coupled by the nonlinearities, one would observe independent dynamics along the $x$ and $y$ directions.

 \begin{figure*}
\includegraphics[width=\textwidth]{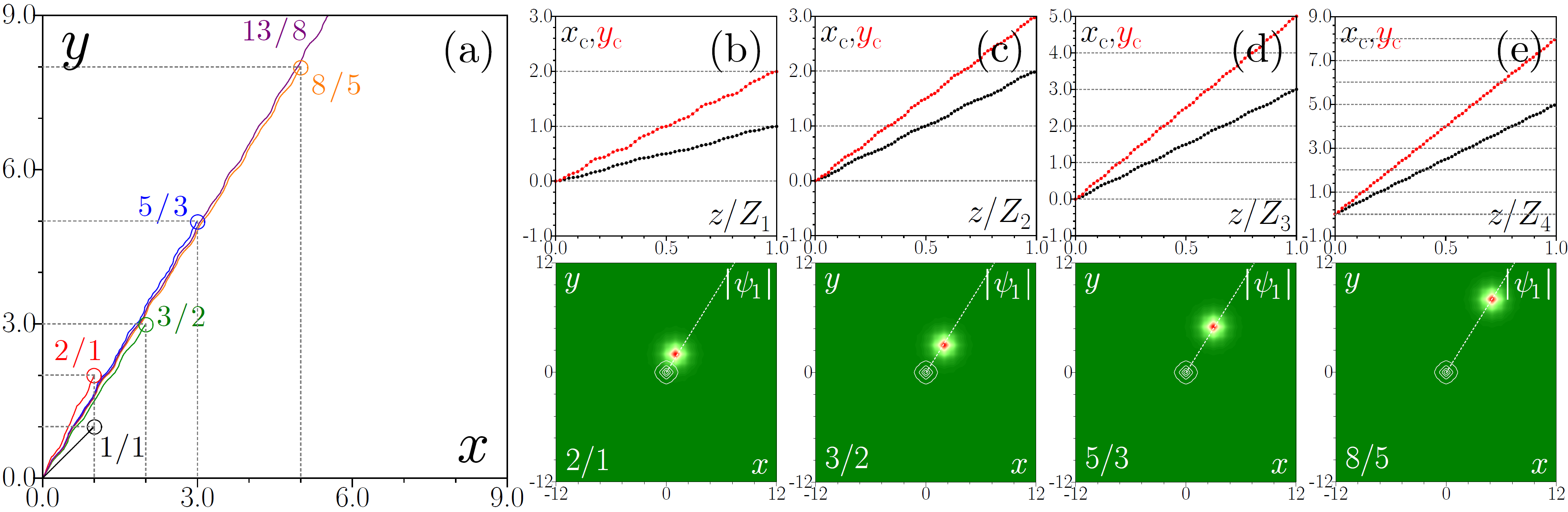}
\caption{\textcolor{red}{Two-dimensional incommensurate bichromatic pumping.} (a) Trajectories of soliton center of mass on the $(x,y)$ plane during one pumping cycle for sliding velocities $\alpha_x=0.01$ and $\alpha_y=(p_n/q_n)\alpha_x$. Input soliton corresponds to $\beta=7$, $\beta=1.01$. BRAs $p_n/q_n$ of the golden ratio $\varphi=(\sqrt{5}+1)/2$ corresponding to each trajectory are indicated on the plot. Circles mark the final soliton positions at $z=Z_n$. Coordinates $x_c$ and $y_c$  of the soliton center of mass versus distance $z$ normalized to one pumping period $Z_n$ for BRA orders $n=1$ (b), $2$ (c), $3$ (d), and $4$ (e). Bottom row in (b)-(e) shows input ($z=0$, while contours) and output ($z=Z_n$, red-green color map) soliton positions. White dashed line corresponds to $y=\varphi x$. Only field modulus distribution in FF component is shown.}
\label{figure5}
\end{figure*}

The dependencies of the $x$ and $y$ one-cycle displacements of the soliton center of mass on propagation constant $b$ for two different BRAs ($n=3$ and $n=4$ orders) of the golden ratio $\varphi$ are shown in Figure~\ref{figure5}. Now output $x_c$ and $y_c$ are obviously different and they increase with increase of the order of approximation $n$. While we again have three different domains of no transport ($b_{\rm co}<b\lesssim 6.4$), transition domain ($6-4\lesssim b\lesssim 6.6$), and the domain where transport is quantized ($b\gtrsim 6.6$), the transition domain manifest rather irregular behavior of the displacement, as compared to the case of commensurate transport shown in Figure~\ref{figure3}. 

\begin{figure}
\includegraphics[width=\columnwidth]{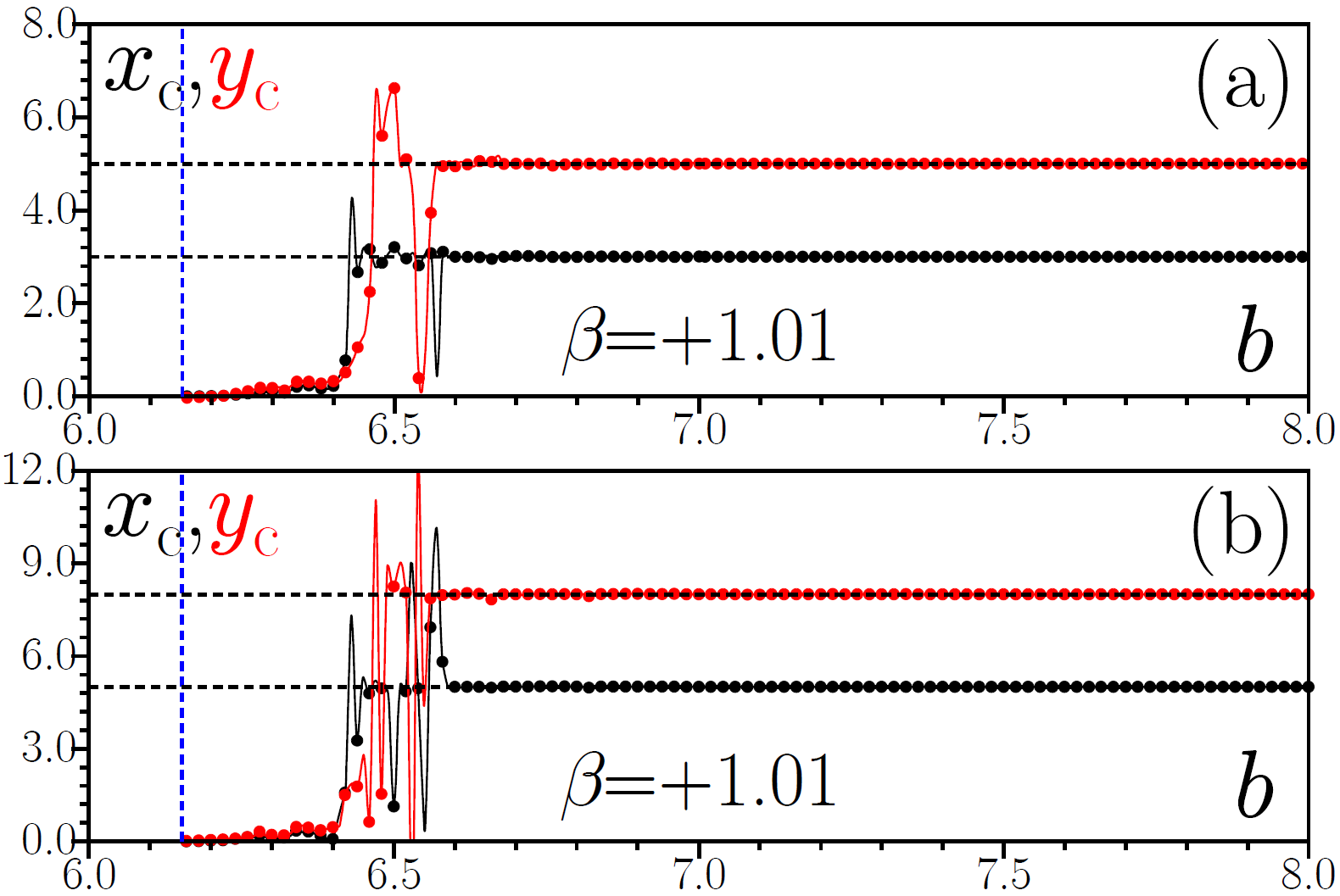}
\caption{\textcolor{red}{Soliton center of mass displacement vs $b$ upon incommensurate bi-chromatic pumping.} Coordinates  $x_c$ and $y_c$ of the the soliton center of mass at versus propagation constant at $\beta=1.01$ for BRAs of the golden ratio of the orders $n=3$ (a) and $n=4$ (b). Sliding velocity is  $\alpha_x=0.01$, $\alpha_y=(p_n/q_n)\alpha_x$. Vertical blue dashed lines indicate cutoff for soliton existence.}
\label{figure6}
\end{figure}

\section{Conclusion}

Summarizing, we have developed theory and illustrated the possibility of stable topological transport of 2D solitons in $\chi^{(2)}$ nonlinear materials. Such transport can occur effectively in any direction in the transverse plane: along one of the lattice axes, in the diagonal direction, or in the direction associated with irrational number depending on the ratio of sliding velocities (and/or periods) of sublattices in the $x$ and $y$ directions. We have found that the transition to stable quantized pumping occurs with increase of soliton's propagation constant, while low-amplitude solitons do not experience pumping either due to instability or due to incomplete band occupation. The predicted pumping is very robust and shows no tendencies for its arrest by the nonlinearity even at very large propagation constants, when soliton contracts effectively to a single lattice period. The dynamics described here show that 2D case offers observation of much reacher pumping scenarios in comparison with 1D case, and most importantly, it allows full control over pumping direction in the transverse plane.







\end{spacing}
\end{document}